\documentclass[prl,twocolumn,superscriptaddress,floatfix]{revtex4-2}
\usepackage{amsmath,amssymb,graphicx}
\usepackage[utf8]{inputenc}
\usepackage[T1]{fontenc}
\usepackage[english]{babel}
\usepackage{url}

\begin{document}
\title{Three-mode chimera-like states in a minimal network of active camphor ribbons}

\author{Ratan \surname{Sarkar}}
\affiliation{Department of Physics, Indian Institute of Technology Bombay, Mumbai 400076, India}

\author{Sapna \surname{Yadav}}
\affiliation{Department of Physics, Indian Institute of Technology Bombay, Mumbai 400076, India}

\author{Punit \surname{Parmananda}}\email{punit@phy.iitb.ac.in}
\affiliation{Department of Physics, Indian Institute of Technology Bombay, Mumbai 400076, India}

\author{Charo I. \surname{del Genio}}
\affiliation{School of Mathematics, North University of China, 030051 Taiyuan, China}
\affiliation{Research Institute of Interdisciplinary Intelligent Science, Ningbo University of Technology, 315104 Ningbo, China}
\affiliation{Bulgarian Academy of Sciences, Institute of Biophysics and Biomedical Engineering, ul. Acad. Georgi Bonchev 21, Sofia 1113, Bulgaria}
\affiliation{International Research Center of Complexity Sciences, Hangzhou International Innovation Institute, Beihang University, Hangzhou 311115, China}
\author{Stefano \surname{Boccaletti}}
\affiliation{International Research Center of Complexity Sciences, Hangzhou International Innovation Institute, Beihang University, Hangzhou 311115, China}
\affiliation{Research Institute of Interdisciplinary Intelligent Science, Ningbo University of Technology, 315104 Ningbo, China}
\affiliation{Sino-Europe Complexity Science Center, North University of China, Taiyuan 030051, China}
\affiliation{Institute of Complex Systems, National Research Council, Sesto Fiorentino 50019, Italy}

\begin{abstract}
Chimera states are a remarkable manifestation of symmetry breaking,
wherein identical units spontaneously split into coherent and incoherent
groups. The term ``chimera'' itself is inspired by the mythological
creature having parts of different animals, reflecting the coexistence
of distinct dynamical behaviors within a single system. Here, we experimentally
demonstrate the existence of a three-mode chimera-like state in a system of four
identical camphor-infused elliptical rotors, each pivoted at a corner
of a square base. When placed on a water surface, the rotors self-propel
and rotate due to Marangoni-driven flows. Chemical coupling mediated
through the shared water interface gives rise to three distinct collective
dynamical states, namely a synchronized one, an antisynchronized one
and an incoherent one. Our results validate recent theoretical work
about three-domain chimeras in reaction-diffusion models, providing
new insights into nonlinear dynamics and suggesting that this new type
of chimera may be observed in a broad class of physical systems.
\end{abstract}

\maketitle

Collective phenomena in networks of coupled oscillators
have broad implications in
disciplines such as physics, biology, chemistry and engineering. Examples
include synchronization~\cite{Win67,Mir90,dewald1991,nair1999mathematical,Pik01,Boc02,Ace05,Boc08,del15,del16,Kum17,Boc18,del22,Jaf24,Par25,Kov26},
clustering~\cite{Gol92,Han93,Oku93,Qi25}, pattern formation~\cite{Kap85,Tay17,Sem25},
amplitude death~\cite{Ata03,dasgupta10,Ban14}, swarming~\cite{van15,OKe17} and
other modes caused by symmetry breaking~\cite{Kel82,Pre15}. Among these,
chimera states are a unique case where oscillators spontaneously split
into a coherent group and an incoherent one. This behavior was first
observed in systems of non-locally coupled phase oscillators~\cite{Kur02},
and, shortly afterwards, named ``chimera'' after the mythological monster,
composed of heterogeneous parts of different animals~\cite{Abr04}. The
first exact results on stability, dynamics and bifurcations of chimeras
were obtained through a minimal model of two interacting oscillator
populations, providing an analytical foundation for the phenomenon~\cite{Abr08}.
Subsequently, numerous studies have deepened our understanding of chimera
states, investigating them across different coupling schemes, topologies
and dynamical regimes~\cite{Wol11,Set14,Ash15,Pan15,Kem16,Mee16,Ber17,Ver15,Ver13,Maj19,Sin20,Par21,Sun25,Guo26}.

A major motivation for studying chimera states comes from biological systems,
where partial synchrony plays a crucial role. For instance, neural systems often
exhibit coexisting coherent and incoherent activity, such as during unihemispheric
sleep in birds and aquatic mammals, or during epileptic seizures, which occur
when a group of neurons suddenly fire in synchrony while others remain desynchronized~\cite{Kim09,Bai12}.
Additionally, chimera states also arise in systems of chemical oscillators~\cite{Tin12,Nko13},
mechanical networks~\cite{Mar13,wojewoda2016smallest} and electronic circuits~\cite{Har16}. However,
despite the large amount of work on the topic, studies of chimeras have so far
usually considered the coexistence of only two states, one synchronized and the
other incoherent.

In this Letter, we report the experimental realization of a chimera-like states consisting
of three modes: synchronized, incoherent and antisynchronized. This three-state
coexistence is reminiscent of the original mythological chimera, a creature composed
of three different animals and typically depicted with the body of a lion, the
head of a goat and the tail of snake. Notably, our experiments
confirm the possibility of the existence of such chimeras, which was introduced
in the theoretical work of Ref.~\cite{Sun25}. In fact, our results extend the
possibilities considered there, since our setup is not a reaction-diffusion system
like the models studied in Ref.~\cite{Sun25}, which feature chimeras states characterized
by the coexistence of more than two spatiotemporal solutions. Although the theoretical model in Ref.~\cite{Sun25} considers a system with infinitely many degrees of freedom, our experiments demonstrate three-mode chimera-like states in a minimal network of only four coupled camphor rotors.

For our experiments, we used camphor-based self-propelling active matter,
where motion is caused by gradients of surface-tension created by camphor
molecules at the air-water interface, which generate self-propulsion via
the Marangoni effect~\cite{Nak97,Son23}. Camphor-based active matter has
long served as an experimental platform for exploring spontaneous propulsion
and nonlinear dynamics at fluid interfaces~\cite{Sha19,Sha20,Bis20,Tiw20,Sha22,Jai23,Tiw23,Mon24,Kha24,dey25},
and classic two-mode chimera-like states have been reported in a system of camphor-infused
rectangular rotors placed on a water surface in an equilateral-triangle configuration~\cite{Sha21}.

Here, our experimental system consisted of four elliptical paper rotors infused
with camphor solution. It needs to be pointed out that, although the rotors were nominally identical,
microscopic variations may have been present because of imperfections in the cutting
process and stochastic fluctuations in camphor deposition.

\begin{figure}[t]
 \centering
 \includegraphics[width=\columnwidth]{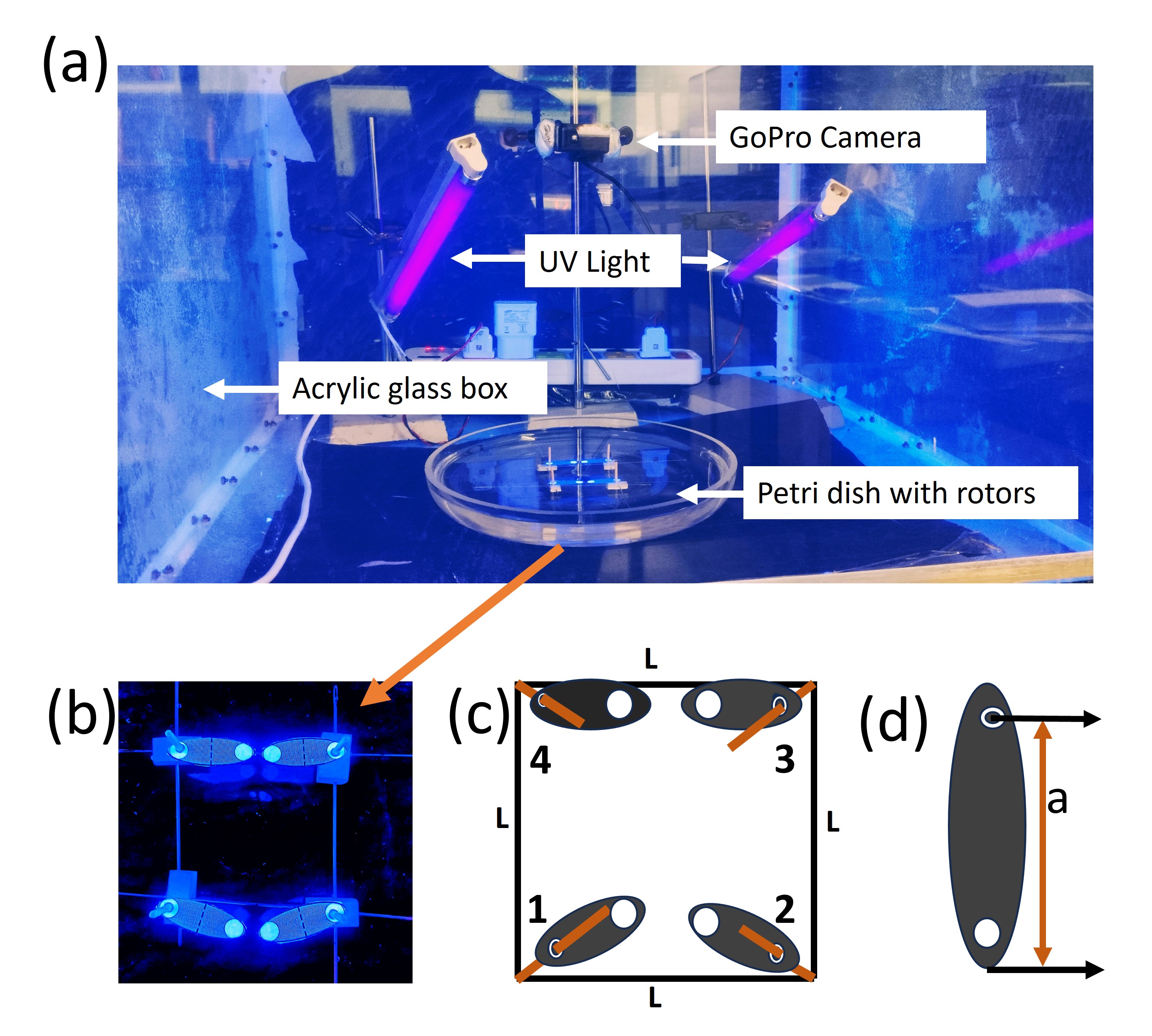}
 \caption{\textbf{Description of the experiment.} (a) Experimental setup. A circular Petri dish filled with deionized (DI) water is placed inside an acrylic enclosure. A square base is positioned at the center of the Petri dish, with camphor-infused elliptical rotors pivoted at each corner of the square. Each rotor is infused with a 2 M camphor–ethanol solution before being placed at the corners of the square base. The elliptical rotors float on the water surface and rotate due to the Marangoni effect. The dynamics of the rotors are recorded from the top using a Go-Pro camera. (b) View from above of the elliptical rotors arranged at the corners of a square base under UV light. The rotors are black, with circular white spots at the end  that reflect the UV light and aid particle tracking. (c) Schematic diagram
 of the square base of arm length $L=5.5$~cm, with the elliptical
 rotors at its corners, labeled~1, 2, 3 and~4. (d) Schematic drawing
 of an individual rotor with pivot to end length $a=2.6$~cm.}\label{fig1}
\end{figure}
In our setup, shown in Fig.~\ref{fig1}, the major axis
of the elliptical rotors was 3~cm in length, whereas the
minor axis was 1~cm. The rotors were entirely black, except
for a white circular spot placed $0.4$~cm from one of
the ends along the major axis to aid visualization. Holes
punched symmetrically to the white spot at $0.4$~cm from
the other end of each rotor, served as pivot points. The
distance from either point to the opposite ends was thus
$2.6$~cm. Prior to the assembly, each rotor was fully
dipped in a 2~M camphor-ethanol solution for 40~s and
then allowed to dry in open air for 180~s. This allowed
the ethanol to evaporate, leaving the rotors coated with
camphor molecules. A square base with side length of $5.5$~cm
was then formed using four needles and four rubber pieces.
Finally, the rotors were mounted with the pivot points
on the corners of the square base, which was placed within
a circular Petri dish of 26~cm of diameter filled with
deionized water, so that the rotors lied on the water
surface. This setup ensured that the rotors would not
physically collide during rotation.

As the rotors come in contact with the water,
they release camphor molecules that locally reduce
the surface tension. Because one side is pinned,
the resulting surface-tension gradient causes
the rotor to start rotating. During the experimental
runs, the setup was illuminated with ultraviolet~(UV)
light, which reflected off the white spots on
the rotors, enabling clear visualization and
tracking of their motion. Video recordings were
captured at 60~frames per second and, to minimize
environmental disturbances such as air currents,
all trials were carried out inside an acrylic
glass enclosure. Room temperature was carefully
maintained at $(23\pm 1)$~°C throughout the experiments.

\begin{figure}[t]
 \centering
 \includegraphics[width=\columnwidth]{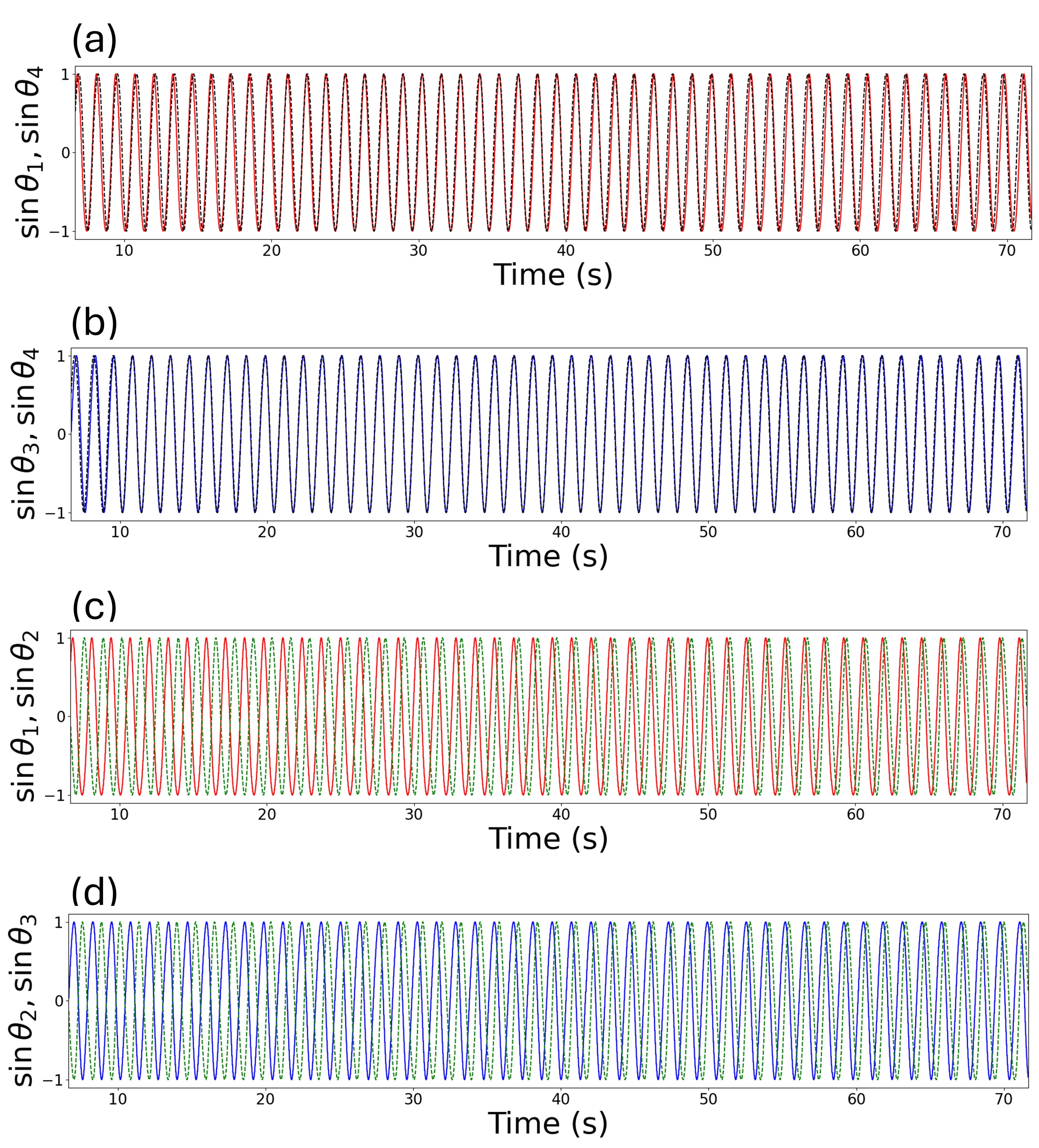}
 \caption{\textbf{Occurrence of three macroscopic dynamical
 states.} The time series of~$\sin\theta_i$ for the four rotors
 after discarding the transients (20~cycles) show that
 the vertical component of the position of rotor~4 (black
 dashed line) is always equal to that of rotor~1 (red solid
 line) and that of rotor~3 (blue solid line). However,
 the vertical component of the position of rotor~2 (green
 dashed line) evolves independently from that of rotor~1
 and that of rotor~3.}\label{fig2}
\end{figure}
Because of their geometric arrangement, the four rotors
can exhibit three macroscopic configurations: (1) all four
rotate in the same direction; (2) three rotors rotate in
the same direction while the remaining one rotates in the
opposite direction; (3) two rotors rotate in the same direction
while the other two rotate in the opposite direction. Note
that the initial direction of rotation of each rotor is
random. In fact, once a rotor starts rotating, it continues
in the same direction until an external perturbation is
applied or its camphor layer is depleted due to diffusion
of camphor molecules into the water. In our experiments,
we observed case~(1) and case~(2), but not case~(3). However,
since the initial conditions are random, we believe that
increasing the number of trials may lead to the observation
of case~(3) as well. Here, we focus on case~(2), where,
even with a minimal network of four rotors, we observed
clear signatures of a chimera-like states, and specifically a
three-state one, in which synchronized, asynchronized,
and antisynchronized dynamics emerged simultaneously within
the system.

\begin{figure}[t]
 \centering
 \includegraphics[width=0.47\textwidth]{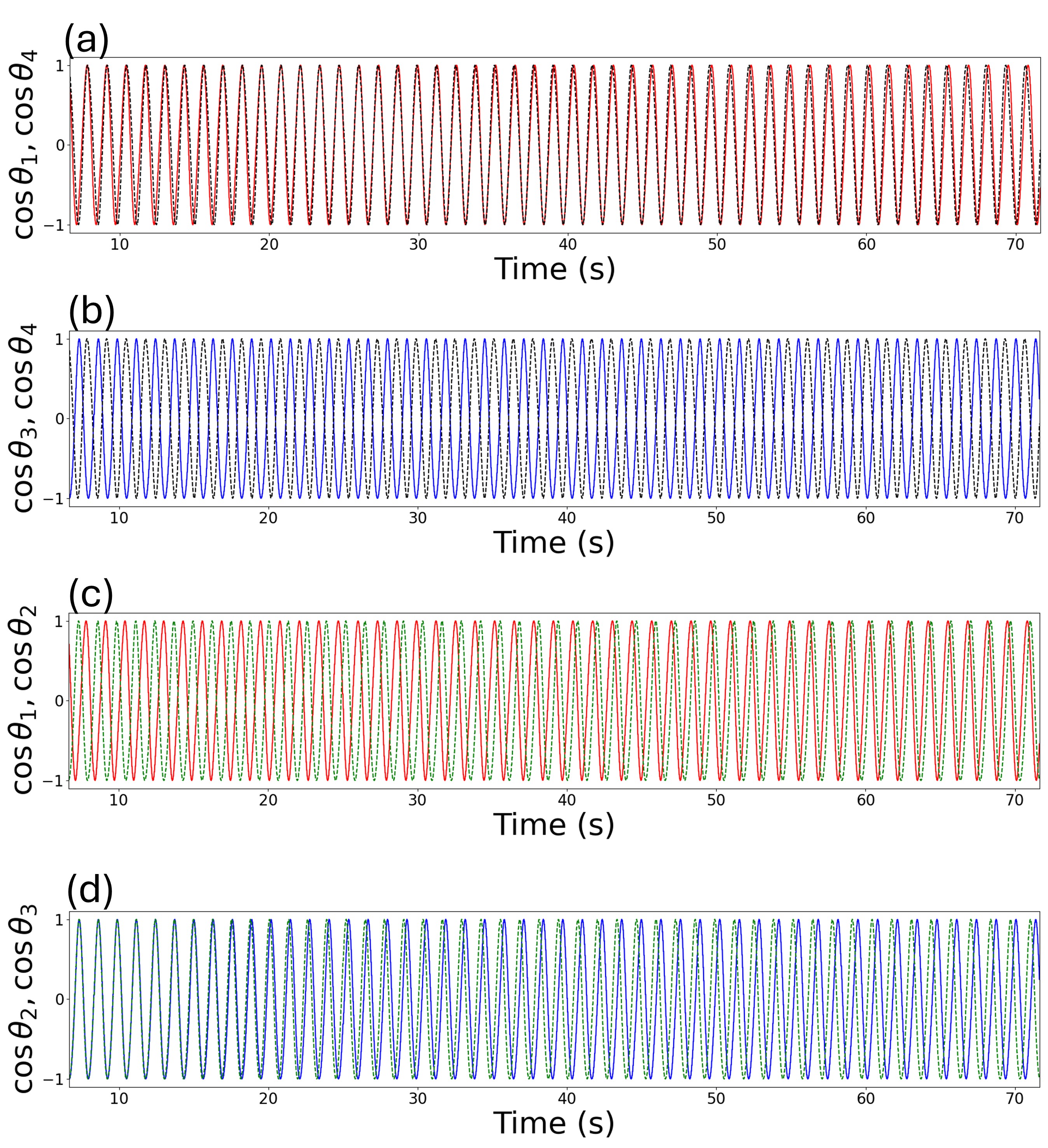}
 \caption{\textbf{Occurrence of three macroscopic dynamical
 states.} The time series of~$\cos\theta_i$ for the four
 rotors after discarding the transients (20~cycles) show
 that the horizontal component of the position of rotor~4
 (black dashed line) is always equal in magnitude and sign
 to that of rotor~1 (red solid line), while being equal
 in magnitude to that of rotor~3 (blue solid line) but
 opposite in sign. Also, the horizontal component of the
 position of rotor~2 (green dashed line) evolves independently
 from that of rotor~1 and that of rotor~3.}\label{fig3}
\end{figure}
\begin{figure}[t]
 \centering
 \includegraphics[width=0.45\textwidth]{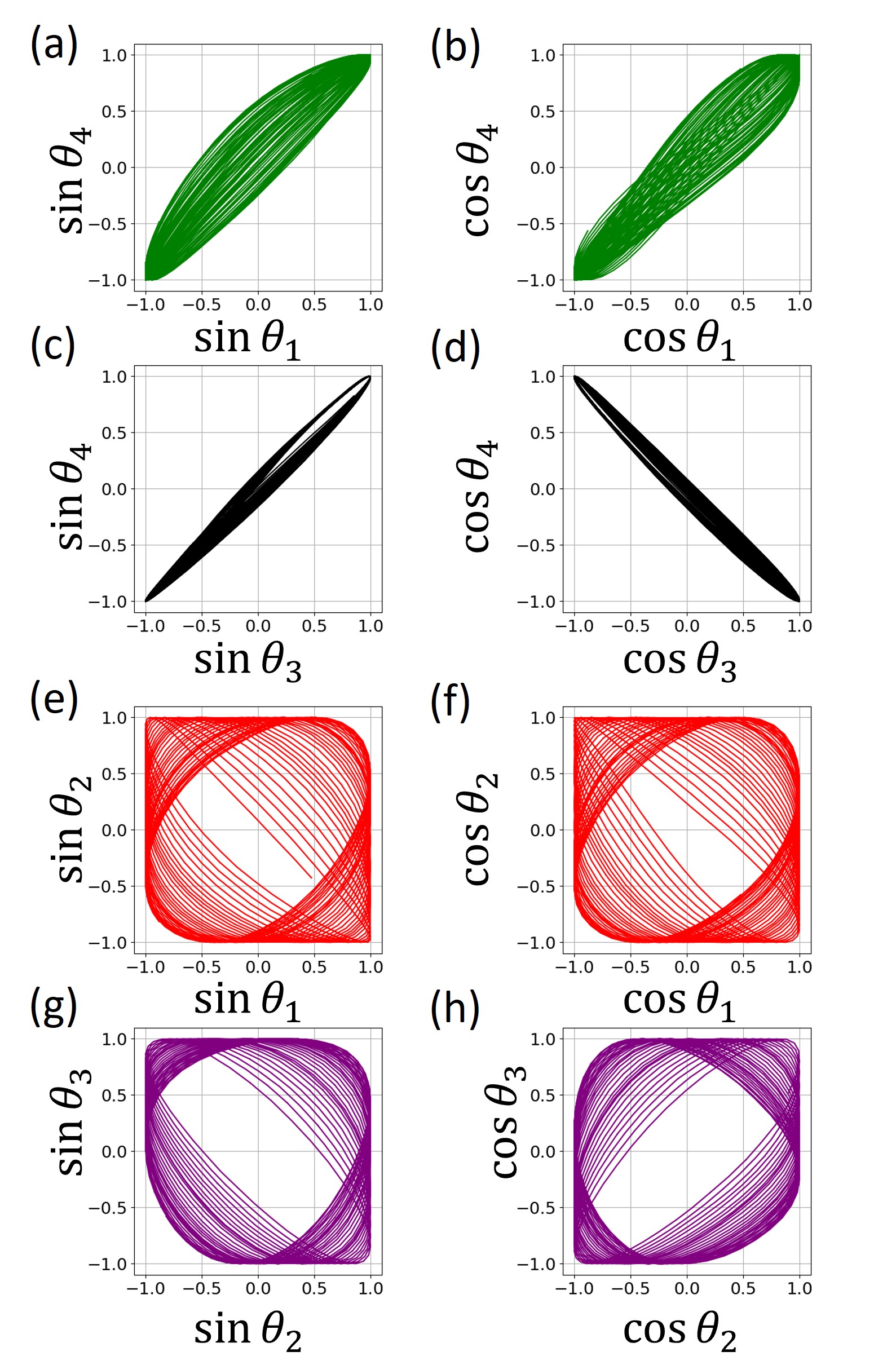}
 \caption{\textbf{Emergence of a three-state chimera.}
 The Lissajous plots of~$\sin\theta_i$ and~$\cos\theta_i$
 for the four camphor rotors confirm the exitence of
 a three-state chimera in our system. Specifically:
 rotor~1 and rotor~4 are synchronized, as both their
 plots (a--b) lie along the bottom-left to top-right
 diagonal; rotor~3 and rotor~4 are antisynchronized,
 as the plot of their vertical components~(c) lies
 along the bottom-left to top-right diagonal, but that
 of their horizontal components~(d) is on the top-left
 to bottom-right one; rotor~2 evolves independently
 from rotor~1 and rotor~3, as the corresponding plots
 occupy the whole space (e)--(h).}\label{fig4}
\end{figure}

Indicating with~$\theta_i$ the angle formed by the $i$th rotor
with the horizontal, the emergence of a three-state chimera can
be identified by studying the temporal evolution of~$\sin\theta_i$
and~$\cos\theta_i$ for all rotors, shown in Fig.~\ref{fig2} and
Fig.~\ref{fig3}, as well as their Lissajous plots, reported in
Fig.~\ref{fig4}. The Lissajous plots offer a particularly intuitive
way of detecting the type of relative motion of two rotors. In
fact, the plots of a pair of synchronized or anti-synchronized
rotors will lie along one of the diagonals, whereas that of a
pair of asynchronous rotors will fill the entire space. More specifically,
in a pair of synchronized elements, both the horizontal components
of the positions and the vertical ones are always equal. Therefore,
the Lissajous figures of the corresponding $\sin\theta_i$ and
$\cos\theta_i$ both lie on the diagonal line from~$(-1,-1)$ to~$(1,1)$.
However, in an antisynchronized horizontal pair of rotors, while
the vertical components are still equal, the horizontal ones are
equal in magnitude but opposite in sign. Then, the Lissajous plot
of the corresponding $\cos\theta_i$ can differentiate synchronized
pairs from antisynchronized ones, since the plot of the former
is still along the same diagonal, whereas that of the latter is
along the diagonal line from~$(-1,1)$ to~$(1,-1)$. Thus, in our
case, we established the following relations: rotors~1 and~4 were
synchronized, as seen from Figs.~\ref{fig2}a, \ref{fig3}a, \ref{fig4}a
and~\ref{fig4}b; rotors~3 and~4 were antisynchronized, as evident
from Figs.~\ref{fig2}b, \ref{fig3}b, \ref{fig4}c and~\ref{fig4}d;
rotors~1 and~2 were asynchronous, as shown in Figs.~\ref{fig2}c,
\ref{fig3}c, \ref{fig4}e and~\ref{fig4}f; rotors~2 and~3 were
also asynchronized, as reported in Figs.~\ref{fig2}d, \ref{fig3}d,
\ref{fig4}g and~\ref{fig4}h. From the Lissajous
patterns in Fig.\ref{fig4}, it is clear that the phase difference
in the case of anti-synchronization is smaller than the one observed
when the rotors are synchronized in-phase. This is because, in
the anti-phase synchronized state, the two rotors come much closer
to each other compared to the in-phase case. This close proximity
leads to stronger interactions and more effective exchange of information
between the rotors. As a result, the phase fluctuations observed
in Fig.~\ref{fig4} for anti-phase synchronization are smaller than
those for in-phase synchronization. For further evidence, see
Supplementary Videos~S1 and~S2 \cite{SI}.

In our experiments, the initial positions
and orientations of the rotors were randomly chosen for each
trial. To ensure reproducibility, we repeated the experiments
multiple times and found that, among the trials leading to
the macroscopic configuration~(2), the  three-mode chimera-like states
were consistently observed, demonstrating the robustness of
the phenomenon, although the specific rotor pairs exhibiting
synchronization, anti-synchronization, or desynchronization
would vary from one run to another. The average lifespan of
the rotors in our experiments is approximately 15--20 minutes.
At the beginning, due to noise in the system, the dynamics
takes some time to reach a stable state. Therefore, we excluded
an initial transient period of about 2--4 minutes from our
analyses. After this transient, once the system becomes stable,
we observed the three-mode chimera-like states for a duration of about
2--5 minutes (approximately~40 to 150~cycles). In principle,
six pairs are possible for the synchronized
and antisynchronized states, namely~(1,2), (1,3), (1,4),
(2,3), (2,4), and~(3,4). However, we did not consider
the diagonal pairs~(1,3) and~(2,4), because these pairs
are not identical to the others in terms of their interaction
distance. Thus, to preserve the requirement of identical
elements in the definition of a chimera-like state, we restricted
our analysis only to the nearest-neighbor pairs~(1,2),
(1,4), (2,3), and~(3,4). Also, the rotors take some time
to reach a steady state. During this transient period,
each rotor rotates with a different angular velocity.
For this reason, we excluded the transients in our analysis.
As time progresses, the rotors interact through the water
surface, as camphor molecules spread across the interface
and create a global coupling. After sufficient time, the
system settles into a stable three-state chimera, in which
one pair becomes synchronized, one becomes antisynchronized,
and the remaining two stay asynchronized.

In conclusion, we experimentally demonstrated  three-mode chimera-like states in a network of four coupled oscillators, characterized by the simultaneous coexistence of synchronized, antisynchronized, and desynchronized groups, wherein the previous work~\cite{Sha21} demonstrated a two-mode chimera-like behavior using a minimal network of three camphor rotors, with synchronized and desynchronized groups coexisting. To the best of our knowledge, the present work provides the first experimental observation of a three-mode Chimera-like state. The interaction between the camphor rotors is mediated through the common air-water interface. Camphor released from each rotor modifies the local surface tension, thereby generating surface-tension gradients that drive the Marangoni flows and influence the motion of the other rotors. Since all four rotors share a common water interface, they are mutually coupled through these camphor-mediated Marangoni interactions, with the strength of the coupling depending on the Euclidean distance between the interacting rotors. By virtue of this mutual coupling, different rotor pairs can exhibit distinct phase relationships coexisting within the observed  collective state. These phase relationships are characterized by nearly constant phase differences for synchronized and antisynchronized pairs, whereas the phase difference continuously varies for desynchronized pairs. The coexistence of these three distinct dynamical relationships gives rise to the observed three-mode chimera-like states. Repeated experiments consistently produced the same  three-mode chimera-like states, highlighting
its robustness. In our experiments, we used
elliptical rotors instead of rectangular ones to minimize water
drag. However, we believe that the  three-mode chimera-like states can also
be achieved with rectangular rotors, as there is nothing that
restricts it to a specific shape.

Additionally, we stress that our work provides
an experimental validation of the possibility of existence of three-mode chimera-like states. These constitute a paradigmatic shift in the study of
collective dynamics, and were investigated thus far only theoretically
in general reaction-diffusion models~\cite{Sun25}. However, it
should also be also noted that our system is not a reaction-diffusion
one, and thus we do not claim that our results are a direct experimental
validation of those in Ref.~\cite{Sun25}. Nonetheless, the very
observation of multiple-domain chimeras in a different type of
system strongly suggests the existence of some more general conditions
for the emergence of these states. Then, since this phenomenon
is observed even in a minimal, physically realizable system, we
expect it to be ubiquitous in nature. This is an exciting insight,
as it indicates that any setup that can support complex broken-symmetry
states can, in principle, constitute a structure in which three-state
chimeras can emerge.

\acknowledgments
RS and SY would like to thank the Indian Institute of Technology Bombay,
India, for financial support. PP acknowledges financial support from SERB,
Department of Science and Technology (DST), India, under Grant No.~CRG/2022/001541).
SB acknowledges support from the Italian Ministry of University and Research under
the PRIN project entitled "The structure, dynamics and control of network
systems with higher-order interaction".

\noindent {Data Availability:}
The data that support the findings of this article and the Supplemental Movies are openly available in~\cite{supplementary}

\end{document}